\documentstyle[prl,aps,twocolumn,psfig]{revtex}

\begin{document}

\twocolumn[\hsize\textwidth\columnwidth\hsize\csname
@twocolumnfalse\endcsname

\draft

\title{Point scatterers in low number of dimensions}
\author{Er'el Granot\footnotemark}

\address{Department of Electrical Engineering, College of Judea and Samaria, Ariel 44837, Israel}

\date{\today}
\maketitle
\begin{abstract}
\begin{quote}
\parbox{16 cm}{\small
It is well known that in 1D the cross section of a point scatterer
increases along with the scatterer's strength (potential). In this
paper we show that this is an exceptional case, and in all the
other cases, where a point defect has a physical meaning, i.e.,
$0<d<1$ and $1<d\leq 2$ ($d$ is the dimensions number), the cross
section does not increase monotonically with the scatterer's
strength. In fact, the cross section exhibits a \emph{resonance
dependence} on the scatterer's strength, and in the singular 2D
case it gets its maximum value for an infinitely weak strength. We
use this fact to show that two \emph{totally different}
generalized functions can describe exactly the \emph{same} physical
entity (the same scatterer).

 }
\end{quote}
\end{abstract}


]

\narrowtext \footnotetext{erel@yosh.ac.il} \noindent It is well
known that in general the cross section of a scattering process
does not have a monotonic dependence on scatterers strength
(potential). On the contrary, the spectral vicinity of resonances
causes the main impact on the scattering process. However, when
small scatterers (compared to the incident scattered particle's
wavelength) are under consideration, no such resonances can occur,
and the scatterer's strength has a direct influence on the
scattering cross section. In 3D this process is known as Rayleigh
scattering in Optics, and Born approximation in Quantum Mechanics
(QM), and unlike Mie scattering (in Optics terminology) has a very
simple dependence on the scatterer's strength: the cross section
is proportional to the square of the scatterer's strength
(scatterer's potential in the QM picture), i.e., $\sigma \propto
V^{2}$. Similarly, in 1D the reflection coefficient monotonically
increases along with the scatterer's strength:
\begin{equation}
R\simeq \left[ 1+4\omega /\left( VL\right) ^{2}\right] ^{-1}
\end{equation}

($L$ is the scatterer's width, $\omega $ is the incoming
particles' energy and $V$ is the scatterer's potential hereinafter
we adopt the units $\hbar =2m=1$).

In this paper we show that these cases are the exceptions: in
general, the cross section does not have a monotonic dependence on
the scatterer's strength, even when the scatterer is merely a {\em
point} defect. In particular we investigate without approximations
the dimensionality regime $0<d\leq 2$ for a point scatterer, and
we show that the $d=1$ is the {\em only case} in this regime, in
which the cross section increases monotonically with the scatterer
strength (potential). In all the other cases (other dimensions),
we prove that beyond a
certain potential (''strength'') value, the scattering cross section starts to {\em %
decrease}.

The stationary-state Schr\"{o}dinger equation in $d$-dimensions
reads:

\begin{equation}
-\nabla ^{2}\psi -p^{2}\psi -\alpha \delta ^{d}\left( {\bf r}\right) \psi =0
\label{shro_eq}
\end{equation}

where we adopt again the units $\hbar =2m=1$ ($m$ is the particle's mass), $%
p^{2}$ is the particle's energy, $\alpha $ is the parameter which
characterizes the point scatterer's strength, $\nabla ^{2}$ is the $d$%
-\emph{dimensions} Laplacian, and $\delta ^{d}\left( {\bf
r}\right) $ is the Dirac delta function in $d$-dimensions
($\bf{r}$ is the vector in $d$-dimensional space).

Let us denote by $G_d\left( {\bf r}\right) $ the $d$-dimensions
Green function, i.e.,

\begin{equation}
-\nabla ^{2}G_d-p^{2}G_d-\delta ^{d}\left( {\bf r}\right) =0
\label{Green}
\end{equation}

and by $\psi _{inc}\left( {\bf r}\right) $ the incident wave
function. It is then straightforward that the scattered wave
function is simply and \emph{exactly}

\begin{equation}
\psi _{sc}\left( {\bf r}\right) =\psi _{inc}\left( {\bf r}\right) +fG\left(
{\bf r}\right)  \label{scatt}
\end{equation}

where the scattering coefficient is equal (exactly) to (for $d<2$)

\begin{equation}
f=\frac{\psi _{inc}\left( 0\right) }{\alpha ^{-1}-G\left(
0\right)}.
\label{scatt_coef}
\end{equation}

Thus, the problem is reduced to an evaluation of the Green
function.

The Green function is required in two regimes: for $\left| {\bf
r}\right| \rightarrow \infty $

\begin{equation}
G_d\left( {\bf r}\right) \sim \left| {\bf r}\right| ^{-\left(
d-1\right) /2}\exp \left( ip\left| {\bf r}\right| \right)
\label{gree_asymp}
\end{equation}

and for $\left| {\bf r}\right| =0$

\begin{equation}
G_d\left( 0\right) =-\frac{\pi ^{\left( 2-d\right) /2}p^{d-2}\exp
\left(
-id\pi /2\right) }{2^{d}\Gamma \left( d/2\right) \sin \left( d\pi /2\right) }%
.  \label{gree_zero}
\end{equation}

($\Gamma$ is the Gamma function, see, for example,
\cite{SpecialFunctions})

Hence,

\begin{equation}
\left| f\right| ^{2}=\frac{\left| \psi _{inc}\left( 0\right) \right| ^{2}}{%
\left\{ \alpha ^{-1}+\left[ \Lambda(d) p^{2-d}\right]
^{-1}\right\} ^{2}+4^{-d}\left( \frac{\pi }{p^{2}}\right)
^{2-d}\Gamma ^{-2}\left( \frac{d}{2}\right) }  \label{scatt_exc}
\end{equation}

where $\Lambda(d) \equiv 2^{d}\pi ^{\left( d-2\right) /2}\tan
\left( \frac{d\pi }{2}\right) \Gamma \left( \frac{d}{2}\right)$.
 We immediately recognize a resonance pattern in
eq.(\ref{scatt_exc}).

In general, the physics encapsulated in Eq.(\ref{scatt_exc}) can
be divided into four categories:

For $0<d<1$, a resonance occurs only for {\em positive} $\alpha $,
i.e., only barriers (in contrast to wells) demonstrate a
non-monotonic dependence on the scaterer's strength ($\alpha $)

For $d=1$, the scattering coefficient (in this case equivalent to
the reflection coefficient) is reduced to the well-known equation
\[
\left| f\right| ^{2}=\frac{\left| \psi _{inc}\left( 0\right) \right| ^{2}}{%
\alpha ^{-2}+\left( 2p\right) ^{-2}}
\]

and the dependence (on $\alpha $) is monotonic (and is independent
of the sign of $\alpha$ ).

For $1<d<2$, the non-monotonic behavior appears again but this
time only for {\em negative} $\alpha $ (that is, only for wells).

$d=2$ is a more complicated case, and will be discussed seperately
in this paper, but the same general nonmonotonic dependence still
holds in the 2D case.

To investigate the resonant nature of Eq.(\ref{scatt_exc}) and the
way it depends on dimensionality, it is more convenient to use,
instead of the enrgy $p^{2}$, the dimensionless energy
\[
\varepsilon \equiv \frac{p^{2}}{\left| \alpha \right| ^{2/\left( 2-d\right) }%
}.
\]

Thus, Eq.(\ref{scatt_exc}) becomes

\begin{equation}
\left| f\right| ^{2}=\frac{\left| \psi _{inc}\left( 0\right) \right|
^{2}\alpha ^{2}}{\cos ^{2}\left( d\pi /2\right) \left[ \left( \epsilon
_{d}/\varepsilon \right) ^{\left( 2-d\right) /2}-1\right] ^{2}+\sin
^{2}\left( d\pi /2\right) }  \label{scatt_exc_s}
\end{equation}

where $\epsilon_d$ is a universal function, which depends merely
on the dimensions number, i.e.,

\begin{equation}
\epsilon _{d}\equiv \pi \left| 2^{d-1}\Gamma \left( d/2\right) \sin \left(
d\pi \right) \right| ^{-2/\left( 2-d\right) }.  \label{eps_d}
\end{equation}

Therefore, the maximum corss section is obtained for
\begin{equation}
\varepsilon =\epsilon _{d}  \label{res_val}
\end{equation}
for any $d$, and the scattering coefficient maximum value is

\begin{equation}
\left| f\right| _{\max }^{2}=\frac{\left| \psi _{inc}\left( 0\right) \right|
^{2}\alpha ^{2}}{\sin ^{2}\left( d\pi /2\right) }  \label{f2_max}
\end{equation}

For the 2D case, Eq.(\ref{f2_max}) may be a bit misleading, for it
suggests an infinitely strong scattering coefficient, i.e.,
$\left| f\right| _{\max }^{2}\sim \left( 2-d\right) ^{-2}$ for
$d\rightarrow 2$. However, this value is obtained, as
Eqs.(\ref{eps_d}) and (\ref{res_val}) imply, only at $\varepsilon
\rightarrow \infty $. For any \emph{finite} energy (i.e.,
$\varepsilon <\infty $) $\left| f\right| ^{2}$ is infinitely small
(like $\sim \left( 2-d\right) ^{2}$).

Before investigating the $d=2$ any further, it is instructive to
investigate the non-monotonic dependence of the scattering
coefficient $\left| f\right| ^{2}$ on the scatterer's strength
$\alpha $. For convenience, let us use the following dimensionless
physical parameters: the scattering coefficient (which is
proportional to the scattering cross section)
\[
{\cal F}\equiv \frac{\left| f\right| ^{2}}{\left| \psi
_{inc}\left( 0\right) \right| ^{2}p^{2 \left( 2-d\right)}}
\]

and the dimensioness strength $\beta \equiv \alpha /p^{\left(
2-d\right) }=\varepsilon ^{(d-2)/2}.$

Then,

\begin{equation}
{\cal F}=\frac{\beta ^{2}}{\cos ^{2}\left( d\pi /2\right) \left[ \beta
\epsilon _{d}^{\left( 2-d\right) /2}-1\right] ^{2}+\sin ^{2}\left( d\pi
/2\right) }  \label{streng_depen}
\end{equation}

\begin{figure}
\psfig{figure=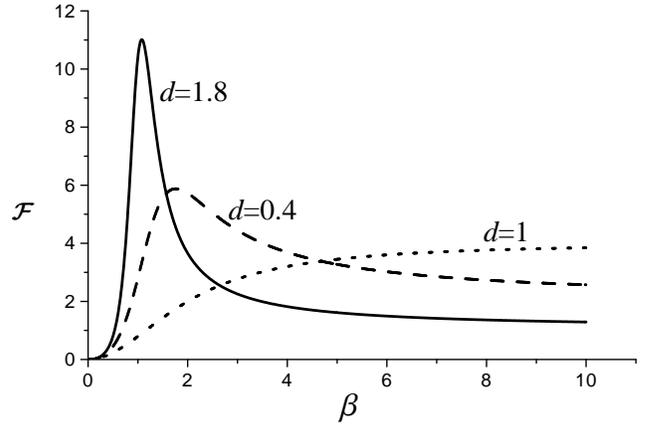,width=10cm,bbllx=45bp,bblly=2bp,bburx=900bp,bbury=530bp,clip=}
\caption{\emph{A plot of the normalized scattering coefficient
${\cal F}$ as a function of the normalized ''strength'' $\beta$
for $d=0.4,1$ and $1.8$. Only in the $d=1$ case there is no local
maximum.}}\label{fig1}
\end{figure}

This function is plotted in Fig.1 for the $d=0.4,1$ and $1.8$
cases. It is clear that in general this function is {\em not}
monotonic. It gets its maximum value for

\[
\beta _{\max }\equiv \varepsilon _{d}^{\left( d-2\right) /2}\sec ^{2}\left(
d\pi /2\right)
\]

for which case

\[
{\cal F}_{\max }\equiv \frac{2^{2d}\Gamma ^{2}\left( d/2\right) }{\pi ^{2-d}}%
,
\]

and for $\beta \rightarrow \infty $,

\[
{\cal F}_{\inf }\equiv \frac{2^{2d}\Gamma ^{2}\left( d/2\right) }{\pi ^{2-d}}%
\sin ^{2}\left( d\pi /2\right) .
\]

and therefore,

\[
\frac{{\cal F}_{\max }}{{\cal F}_{\inf }}=\frac{1}{\sin ^{2}\left( d\pi
/2\right) }.
\]

Two universal dimensions appear: ${\cal F}_{\max}$ receives its
minimum value for $d \simeq 0.81$ and ${\cal F}_{\min}$ receives
its maximum value for $d=1.19$, however they get the same value
with the same derivative \emph{only} at $d=1$ (see Fig.2).

\begin{figure}
\psfig{figure=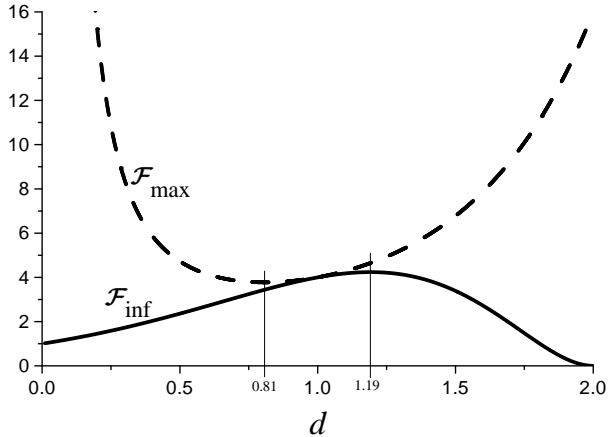,width=10cm,bbllx=50bp,bblly=20bp,bburx=900bp,bbury=540bp,clip=}
\caption{\emph{A plot of ${\cal F}_{\max}$ and ${\cal F}_{\inf}$
as a function of $d$. }}\label{fig2}
\end{figure}

It is then clear why monotonic dependence is restored {\em only}
for $d=1$.
In all the other cases (both $d<1$ and $d>1$), ${\cal F}_{\inf }<{\cal F}%
_{\max }$ (see Fig.1); moreover, the larger the dimensionality
(beyond $d>1$), the lower $\beta _{\max }$. Therefore, to achieve
maximum cross section, the scatterer's strength should be
\emph{decreased}.

These last two conclusions run contrary to common intuition. One
would normally expect to increase the scattering cross section by
increasing the scatterer strength, not decrease it.

With this in mind, it should not be surprising that in order to
get a maximum scattering cross section in 2D the point scatterer's
strength $\alpha$ should be infinitely low, and for any other
finite value of $\alpha$, the cross section will vanish.

The applicability of a point scatterer in 2D is well known in the
literature, and due to the singularity of the 2D case, there where
many indirect approaches to model a 2D point scatterer (see, for
example,
\cite{Azbel,Chu_Sorbello_89,Kumar_91,Bagwell_90,Levinson_Lubin_Sukhorukov_92,Gurvitz_Levinson_93,Tekman_90,Kim_99}).
In what follows we present two generalized function, which can
replace directly the potential of a 2D point scatterer without any
approximations. The first function was already presented by Azbel
\cite{Azbel} , but the second one is a new function. Despite the
different appearance they both describe the same defects.

To create a scatterer (with a finite cross section) in 2D one can
choose, instead of a delta function, a finite width ($\rho$)
scatterer (but a very narrow one, i.e.,$\rho p\ll 1$)
\cite{Azbel}. For example, let us choose the potential
\cite{Azbel}

\begin{equation}
V\left( {\bf r}\right) =-\alpha \frac{\exp \left[ -\left( r/\rho \right) ^{2}%
\right] }{\pi \rho ^{2}}  \label{two_d_pot}
\end{equation}

Surely, in the limit $\rho \rightarrow 0$, eq.\ref{two_d_pot} is
simply the Dirac's delta function.

Since this potential has a finite width, one cannot use
eq.\ref{gree_zero} for the Green function. Instead, the Green
function near $|\textbf{r}|=0$ should be taken, and in 2D it can
be written as

\begin{equation}
G\left( r\right) \simeq -\left( 2\pi \right) ^{-1}\ln
\left(ipre^{\gamma}/2\right), \label{G2D_rsmall}
\end{equation}

where $\gamma $ is the Euler constant \cite{SpecialFunctions}.

Moreover, due to the singularity one cannot use
eq.(\ref{scatt_coef}) to calculate the scattering coefficient,
instead the following general expression should be taken
\cite{Granot_Azbel}

\begin{equation}
f=-\frac{\psi _{inc}\left( 0\right) \int
d\textbf{r}'V(\textbf{r}') }{1+\int
d\textbf{r}'G(\textbf{r}')V(\textbf{r}')}.
\label{scatt_coef_for2D}
\end{equation}

Substituting eqs. \ref{G2D_rsmall} and \ref{two_d_pot} in
eq.\ref{scatt_coef_for2D}, the scattering coefficient will be

\begin{equation}
f\simeq\frac{\psi _{inc}\left( 0\right) }{\alpha ^{-1}+\left( 2\pi
\right) ^{-1}\ln \left( p\rho e^{-\gamma /2}/2\right) +i/4}
\end{equation}

This expression again illustrates the non-monotonic behavior of
$|f|$ on the scatterer's strength $\alpha$: for a very weak
scatterer, $\left| f\right| $ increases with the strength, but
when $\alpha $ exceeds the value $-\left(
2\pi \right) \left[ \ln \left( p\rho e^{-\gamma /2}/2\right) \right] ^{-1},$%
\ $\left| f\right| $ begins to decrease. In particular, choosing
(ref. \cite{Azbel})

\begin{equation}
\alpha =\frac{2\pi }{\ln \left( \rho _{0}/\rho \right) }  \label{alpha-ex}
\end{equation}

one obtains (in the limit $\rho \rightarrow 0$)

\begin{equation}
{\cal F=}\frac{16}{\pi ^{-2}\ln ^{2}\left( p^{2}/E_{0}\right) +1}
\label{result}
\end{equation}

where

\begin{equation}
E_{0}\equiv \frac{4}{\rho _{0}^{2}}\exp \left( \gamma \right)
\label{relation}
\end{equation}

is the point potential resonance energy.

We therefore see again that in 2D, the point impurity potential
must be infinitely shallower than the 2D delta function, that is,
the strength   is infinitely (logarithmically with the impurity's
width) narrow (eq.\ref{alpha-ex}).

In the last few paragraphs we confronted the problematic (due to
the singularity) 2D case by starting with a finite ($\rho$) width
potential, fixing the right potential strength (eq.\ref{alpha-ex})
and only then taking the limit $\rho \rightarrow 0$. More
formally, the desired potential can be written

\begin{equation}
V\left( {\bf r}\right) =-{\lim_{\rho \rightarrow 0} }\frac{2\exp %
\left[ -\left( r/\rho \right) ^{2}\right] }{\rho ^{2}\ln \left(
\rho _{0}/\rho \right) }  \label{IDF}
\end{equation}

Such a generalized point potential mimics a physical 2D point
scatterer with eigen energy expressed by eq. \ref{relation}.

This procedure was taken in \cite{Azbel} by Azbel and was named
Impurity D Function (IDF), however, this is not the only way to
attack the singular 2D potential. Our derivation also suggests
that it is possible to confront the singularity of the 2D case via
the dimensions' number (i.e., $d$). That is, it is clear that
$d=2$ and $\rho=0$ is a singular point, but instead of approaching
it via the $\rho \rightarrow 0$ limit (while fixing $d=2$), one
can approach it via the $d \rightarrow 2$ (from below) limit
(while keeping $\rho=0$).

\begin{figure}
\psfig{figure=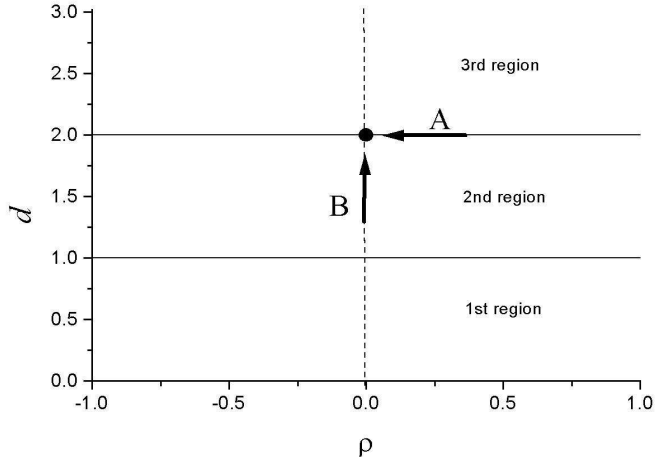,width=10cm,bbllx=50bp,bblly=20bp,bburx=600bp,bbury=400bp,clip=}
\caption{\emph{In the $d-\rho$ plane, each arrow illustrates a
different way of approaching the singular point $d=2$ and $\rho=0$
. The 1st and 2nd regions (including the $d=2$ line) correspond to
impurities whose cross section does not depend monotonically on
their potential strength. The $d=1$ line is the only case in which
the monotone dependence occurs. In the 3rd region, a point
impurity cannot exist. Notice that the $\rho<0$ region has no
physical meaning.}}\label{fig3}
\end{figure}

In particular, again choosing the potential

\[
V\left( {\bf r}\right) =-\alpha \delta ^{d}\left( {\bf r}\right)
\]

with the potential ''strength''

\[
\alpha =-2\pi \left( 2-d\right) E_{0}^{\left( 2-d\right) /2}
\]

substituting it in Eq.(\ref{streng_depen}) and then taking the limit $%
d\rightarrow 2$ (from below), the simple expression (\ref{result})
of ${\cal F}$ is retrieved.

We therefore find another way to construct a 2D point impurity,
which describes a physical impurity characterized by the eigen
energy  $E_{0}$. This generalized potential can be written as the
following limit

\begin{equation}
V\left( {\bf r}\right) =-2\pi {\lim_{d\rightarrow 2}}(2-d)
E_{0}^{\left( 2-d\right) /2}\delta ^{d}\left( {\bf r}\right) .
\label{potential}
\end{equation}

The potentials described by eqs. \ref{IDF} and \ref{potential}
describe \emph{exactly} the same physical entity. Despite their
different appearance, they both present a point impurity in 2D,
with the same eigen energy $E_{0}$ (see eq.\ref{relation}).

In both cases, these are generalized functions, where their limit
is presented in Fig.3: the limit $\rho \rightarrow 0$ of eq.
\ref{IDF} is presented by the "A" arrow in the figure, while the
limit $d \rightarrow 2$ of eq.\ref{potential} is the "B" arrow.

To summarize, in this paper we showed that our everyday experience
that the cross section of a point scatterer increases with the
scatterer's strength fails in low number of dimensions. It was
shown that the 1D is the only case in which there is a monotonic
dependence of the scattering coefficients on the scatterer's
strength, in all the other cases (other dimensions) the scattering
coefficient shows a resonance dependence on the scatterer's
strength. We use our derivation to construct a new generalized
function, which can describe the potential of a point defect in
2D.

\end{document}